# A quantum plasmonic nanocircuit on a semiconductor platform

Xiaofei Wu[1,2,3,4], Ping Jiang[1,5], Gary Razinskas[3,4], Yongheng Huo[6,8], Hongyi Zhang[2,9], Martin Kamp[4], Armando Rastelli[6,10], Oliver G. Schmidt[6], Bert Hecht[3,4], Klas Lindfors[2,7]* and Markus Lippitz[1,2]*

[1]Experimental Physics III, University of Bayreuth, Universitätsstraße 30, 95447 Bayreuth, Germany

[2]Max-Planck-Institute for Solid State Research, Heisenbergstraße 1, 70569 Stuttgart, Germany

[3]Experimental Physics V, University of Würzburg, Am Hubland, 97074 Würzburg, Germany

[4]Wilhelm Conrad Röntgen Research Center for Complex Material Systems, University of Würzburg, Am Hubland, 97074 Würzburg, Germany

[5]College of Science, China University of Petroleum, Changjiang West Road 66, Qingdao 266580, China

[6]Institute for Integrative Nanosciences, IFW Dresden, Helmholtzstraße 20, 01069 Dresden, Germany

[7]Department of Chemistry, University of Cologne, Luxemburger Straße 116, 50939 Köln, Germany

[8]Hefei National Laboratory for Physical Sciences at Microscale and Department of Modern Physics, Shanghai Branch, University of Science and Technology of China, Shanghai 201315, China

[9]Center for Quantum Information, Institute for Interdisciplinary Information Sciences, Tsinghua University, Beijing 100084, China

[10]Institute of Semiconductor and Solid State Physics, Johannes Kepler University Linz, Altenbergerstraße 69, 4040 Linz, Austria

*e-mail: klas.lindfors@uni-koeln.de; markus.lippitz@uni-bayreuth.de

**Quantum photonics holds great promise for future technologies such as secure communication, quantum computation, quantum simulation, and quantum metrology[1]. An outstanding challenge for quantum photonics is to develop scalable miniature circuits that integrate single-photon sources, linear optical components, and detectors on a chip. Plasmonic nanocircuits will play essential roles in such developments[2–4]. Plasmonic components feature ultracompact geometries and can be controlled more flexibly and more energy-efficiently compared to conventional dielectric components due to strong field confinement and enhancement[5,6]. Moreover, plasmonic components are compatible with electronic circuits, thanks to their deep subwavelength sizes as well as their electrically conducting materials[7–9]. However, for quantum plasmonic circuits, integration of stable, bright, and narrow-band single photon sources in the structure has so far not been reported. Here we present a quantum plasmonic nanocircuit driven by a self-assembled GaAs quantum dot. The quantum dot efficiently excites narrow-band single plasmons that are guided in a two-wire transmission line until they are converted into single photons by an optical antenna. Our work demonstrates the feasibility of fully on-chip plasmonic nanocircuits for quantum optical applications.**

Self-assembled semiconductor quantum dots are widely used single photon sources as they are stable, non-blinking, electrically drivable, and have well-defined orientations of the transition dipole-moments[10–12]. They have been also integrated in dielectric on-chip devices by structuring the semiconductor material[13]. However, the high refractive index of the semiconductor host material makes it challenging to couple self-assembled quantum dots with surface plasmons. Therefore, in quantum optics experiments with plasmonic circuits, other single emitters such as colloidal nanocrystals and nitrogen vacancy centers in diamond nanocrystals[2,14], or nonlinear optical processes[3,4] were used as single photon sources. For the former case, however, these emitters suffer from bleaching, blinking, and poor control of the orientation of the transition dipole moment. Their emission is also often more broadband than that of self-assembled quantum dots. Finally, these emitters are not easily excited electrically and are hardly suited for scalable devices. Nonlinear optical methods meanwhile are hindered by the bulky sources and low photon rates. It is thus of great interest to develop methods to integrate self-assembled semiconductor quantum dots as single-photon sources in plasmonic nanocircuits.

We address the problem posed by the high refractive index of self-assembled quantum dot samples by releasing the emitters from the bulk crystal and exploiting an indirect-coupling approach. As illustrated in Fig. 1a, the structure consists of a bar of AlGaAs heterostructure, which contains a GaAs quantum dot, and a pair of gold wires on both sides of the AlGaAs bar. The structure is placed on a SiO$_2$ substrate and constitutes an in-plane version of a hybrid waveguide[15]. The transition dipole moments of the two near-degenerate excitons in the quantum dots are in the plane of the substrate. The hybrid waveguide is transformed into a plasmonic waveguide when the semiconductor bar is tapered, leaving behind only the gold wires. The electromagnetic energy in the hybrid waveguide is thus coupled into surface plasmons propagating along the gold-wire transmission line (Fig. 1b).

Numerical analysis reveals that the coupling between the quantum dot and the hybrid waveguide is dominated by the transversal component (y-direction in Fig. 1b) of the transition dipoles and one of the hybrid waveguide modes (Fig. 1c, see also Supplementary Fig. S2-S4 for details of the modes). In this hybrid mode, the electric field component along the y-axis $E_y$ is much stronger than the other two components and is concentrated at the center of the semiconductor. 44 % of the power emitted by a transversal dipole source at the center of the waveguide is transmitted into each propagation direction of the hybrid mode. When the field propagates through the taper, the hybrid mode evolves into a plasmonic mode of the gold-wire transmission line (Fig. 1c) with a conversion efficiency of 58%. Taking into account the propagation loss of the hybrid mode, we calculate the plasmon excitation efficiency for the dipole position shown in Fig. 1b as 25 %. A quantum dot in a bulk, unprocessed crystal, which is used in traditional quantum optical experiments as single-photon source, emits only 2% of its power into free space, and even less is collected by a microscope objective. Our device thus functions as a bright single-plasmon source for on-chip quantum optics. The overall efficiency could be significantly improved by adding a reflector to the hybrid waveguide section to obtain constructive interference in unidirectional propagation. This would also result in an increased radiative decay rate of the quantum dot. In the present design, the plasmonic structures do not modify the decay rate of the dot as they are rather far away.

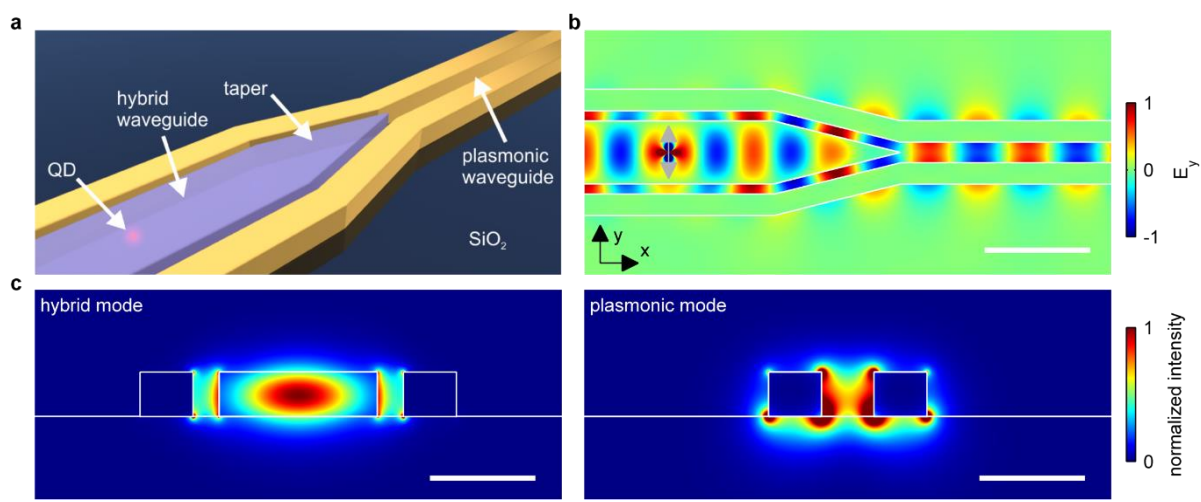

**Figure 1 | Geometry of the structure and numerically simulated properties. a,** Schematic of the structure for efficient excitation of a plasmonic waveguide with a self-assembled quantum dot. **b,** Instant distribution of $E_y$ in the midplane of the structure when excited with a y-polarized dipole (light gray arrow) at the center of the waveguide, showing the propagating wave and mode conversion at the taper. The scale bar is 500 nm. **c,** Profiles of the relevant modes of the hybrid waveguide (left) and the plasmonic waveguide (right). The electric-field intensity of each panel is normalized in such a way that both modes have the same power. The scale bars are 200 nm. The outlines of the structure are overlaid in **b** and **c**. Detailed dimensions are given in Supplementary Fig. S1.

We prove our idea in experiments with a simple plasmonic circuit model. A scanning electron micrograph of a fabricated device is shown in Fig. 2a. The AlGaAs bar is tapered at both ends and the gold wires are terminated with optical antennas. The optical antennas have been designed to be resonant at the quantum dot emission wavelength of approximately 780 nm. When exciting the quantum dot in the AlGaAs bar with a stationary focus of a continuous-wave laser (532 nm wavelength), a raster-scanned confocal detection focus finds bright and stable emission at the position of both the quantum dot and the antenna on the right, shown in Fig. 2b for y-polarized emission. Note that the antenna emission is brighter than the direct radiation of the quantum dot, suggesting that plasmons are efficiently launched in the gold-wire transmission line. This experimental result agrees with the simulated far-field image (Fig. 2c, see also Supplementary Fig. S6). In the simulation, the integrated intensity ratio of the antenna spot to the quantum dot spot is larger than in the experiment. We attribute this disagreement to imperfections of the fabricated structure. Particularly, unexpected loss in the hybrid waveguide is evidenced by additional experimental results (Supplementary Fig. S7) and accounts for the dark left antenna in Fig. 2b.

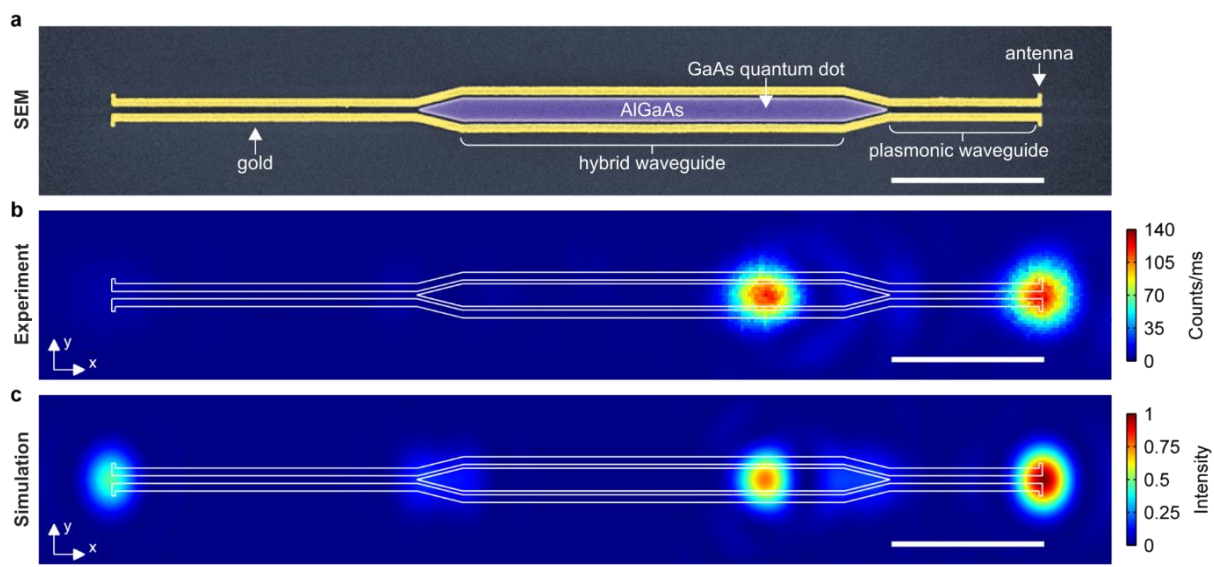

**Figure 2 | Experimental nanocircuit and observation of efficient plasmon excitation. a,** Scanning electron micrograph of the fabricated structure. A GaAs quantum dot in the AlGaAs bar couples with the hybrid waveguide and excites the plasmonic waveguides through the taper. The plasmons are then guided by the gold-wire transmission lines to the optical antennas. **b,** Photoluminescence micrograph of the structure. The quantum dot is excited by a stationary laser focus. The y-polarized emission is collected by raster-scanning a confocal detection focus. **c,** Numerically simulated far-field image following the experimental structure and conditions closely. The excitation source is a pair of incoherent x and y-polarized dipoles at the quantum dot position of panel **b,** and only $|E_y|^2$ in the image plane is plotted. The outlines of the structure are overlaid in **b** and **c**. All scale bars are 2 μm. Detailed dimensions are given in Supplementary Figs. S1 and S5.

To assess the suitability of our plasmon source for quantum optical experiments, we now turn to the spectral and statistical properties of the emitted light. Photoluminescence spectra taken from the spots at the right antenna and the quantum dot of Fig. 2b exhibit exactly the same features in terms of the spectral position and width of the sharp lines and their relative intensities (Fig. 3a). As these features are a finger print for each quantum dot, the spectral correlation shown here further confirms that the quantum dot in the structure is the source of the emission of the antenna. Most importantly, the exciton line is rather narrow for a near-surface quantum dot, although not resolution-limited. We suspect that the sample fabrication has caused broadening of the exciton line, as unprocessed quantum dots show a resolution-limited linewidth of 60 μeV.

To verify that the quantum statistics of the quantum dot emission is preserved by our device and that we really launch single plasmons, we measured the second-order cross-correlation function $g^{(2)}(\tau)$ between the emission in the exciton line collected at the antenna and the quantum dot (Fig. 3b). The antibunching dip at time delay $\tau$ = 0 goes well below 0.5, indicating that the quantum dot emits single photons that are coupled into single plasmons in the gold-wire transmission line, and the plasmons are subsequently converted into single photons again by the antenna. At saturated excitation, the detected photon rate is 2 kcps in the exciton line. Taking the detection efficiency of the setup and the simulated collection efficiency into account, we estimate a single-plasmon rate of $3\times10^6$ s$^{-1}$ at the taper-end. The ability to efficiently excite and operate with single plasmons provides the basis for quantum applications.

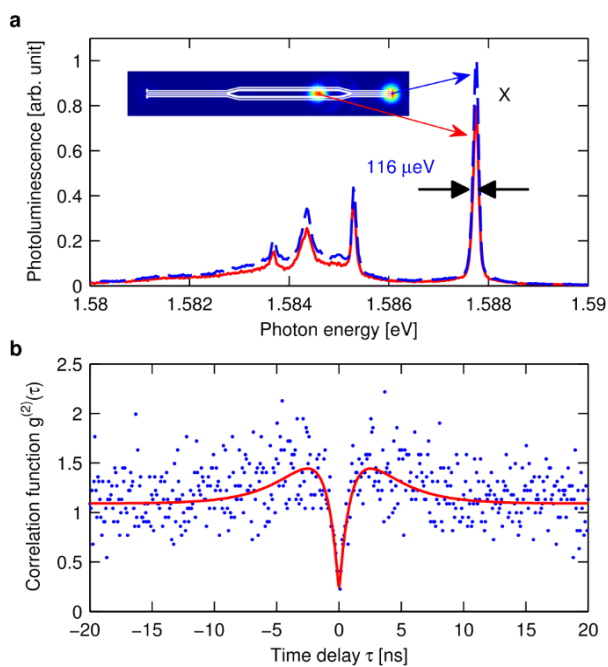

**Figure 3 | Narrow bandwidth single plasmons. a,** Spectra collected at the position of the quantum dot (red solid line) and antenna at the right end of the structure (blue dashed line). The neutral exciton peaks (X) of both spectra have the same full width at half maximum of 116 µeV. **b,** Measured second-order cross-correlation function $g^{(2)}(\tau)$ between the antenna and quantum dot (blue dots). The red line is a fit to model[30], taking the off-resonant excitation conditions into account.

We have demonstrated a simple model quantum plasmonic nanocircuit with a narrow-band self-assembled GaAs quantum dot as a source for single plasmon excitation. Based on our approach, more complex plasmonic as well as photonic circuits can be developed (Supplementary Fig. S8). For example, multiple quantum dots can be coupled via plasmonic circuits to achieve photonic transistors[16]. By applying voltages on the gold wires, the transitions of the quantum dots can be electrically tuned to be resonant. By using site-controlled self-assembled quantum dots, additional flexibility with device design can be obtained[13,17]. *In situ* lithography approaches using photoluminescence[18] or cathodoluminescence[19] could also help to position quantum dots in devices. Plasmonic components such as interferometers, modulators, and switches[6] can be readily integrated due to the flexible electron-beam lithography (EBL) based fabrication. It is also possible to convert the plasmonic mode in Figure 1c to the other mode of the gold-wire transmission line[20] or to other waveguide modes (e.g. modes on a single gold wire) to fit different purposes. Moreover, electrical excitation of the quantum dot[21,22] and electrical detection of single-plasmons[3,23,24] can be implemented to make all-on-chip circuits.

Semiconductors are well-developed materials for integrated electronic circuits and a wide variety of semiconductor-based quantum photonic circuits are being established as well[25–27,13]. Our work shows that by making use of the self-assembled quantum dots, quantum plasmonic circuits can also be built on semiconductor platforms, and it thus opens the way to integrate electronic, photonic, and plasmonic devices on one semiconductor chip for applications of quantum technologies.

**Methods**

**Numerical simulations.** The mode profiles were obtained using a full-vectorial eigenmode solver (MODE Solutions, Lumerical Solutions Inc.). The three-dimensional (3-D) electromagnetic simulations of the structures were performed using the finite-difference time-domain method (FDTD Solutions, Lumerical Solutions Inc.). The coupling efficiencies into waveguide modes were calculated by means of a mode expansion analysis, calculating the overlap integrals of the field distribution in 3-D structures and the previously obtained mode profiles. Far-field images were obtained by projection of the simulated near-field intensity distribution recorded 10 nm above the structure into the air half-space taking the numerical aperture (NA) of the collection objective into account. All simulation results are for vacuum wavelength of 780 nm.

**Materials and sample fabrications.** The GaAs quantum dots in AlGaAs barriers were grown by molecular beam epitaxy on a GaAs substrate as described in detail in Refs. [28,29]. A sketch of the quantum dot structure is shown in Supplementary Fig. S5. The epitaxial layers were released from the GaAs substrate by etching the sacrificial AlAs layer. The released membrane was transferred to a substrate of 400 nm $SiO_2$ on Si. The thickness of the $SiO_2$ is chosen to find a compromise in minimizing charging effect for EBL and the influence of Si on the optical properties of the waveguide, and having a constructive interference between the direct emission and the reflection of the $SiO_2$-Si interface. The membrane was then etched into arrays of tapered bars using EBL and reactive ion etch. A 3 nm thick $Al_2O_3$ layer was deposited on the sample immediately after etching using atomic layer deposition to protect the semiconductor from oxidation and further processing. Finally, the gold wires and optical antennas for each structure were created by the second step of EBL and thermal evaporation deposition. Alignment between the semiconductor and metal structures was achieved using markers patterned in the first lithography step. See Supplementary Fig. S5 for more details about the sample.

**Optical measurements and data processing.** A sketch of the optical setup is shown in Supplementary Figure S9. A 532 nm continuous-wave laser diode was used as the excitation source. The laser was focused on the sample by an objective of NA 0.7 through the cryostat window. The sample was cooled down to 10 K temperature. For the confocal luminescence image, a single photon avalanche diode (SPAD) was used to detect the emission in the spectral band of 773 – 792 nm. A monochromator and a CCD camera were used for acquiring the spectra. The waveguide image was aligned parallel to the grating lines. For the correlation function measurements, the spots of the quantum dot and the antenna were dispersed by the monochromator and only the photons corresponding to the exciton line were transmitted to the two SPADs. The timing output signal of the SPADs were sent to a time-correlated single photon counting unit operating in time-tagged time-resolved mode. The fitting model takes the off-resonant excitation into account[30] (see Supplementary Method for more details). The optical characterization results in Figs. 2 and 3 are from the same structure. We have determined that the quantum dot investigated here resides exactly at the center of the semiconductor bar in the transversal direction (Supplementary Fig. S10), providing an ideal structure for the experiments.

**Acknowledgments**

We gratefully acknowledge financial support from the Deutsche Forschungsgemeinschaft through SPP1391 Ultrafast Nanooptics. K.L. thanks the University of Cologne for funding through the Institutional Strategy of the University of Cologne within the German Excellence Initiative and the support of the Academy of Finland (Project No. 252421). P.J. acknowledges financial support from the National Natural Science Foundation of China (No. 11204381) and China Scholarship Council. We additionally would like to thank Jürgen Weis and the Nanostructuring Lab team of the Max Planck Institute for Solid State Research for help with sample fabrication and Christian Dicken for help with optical measurements. X.W. thanks Jianjun Chen for valuable discussions.

**Author contributions**

K.L. and M.L. conceived the project. X.W. designed the structures and experiments and constructed the optical setup. Y.H. grew the GaAs quantum dots under supervision of A.R. and O.G.S. X.W. and H.Z. developed the quantum dot membrane transfer method. X.W. and K.L. fabricated the samples. M.K. prepared the transmission electron microscopy lamella and performed the TEM characterization. X.W., P.J. and M.L. carried out the optical measurements and analyzed the data. G.R. performed simulations under supervision of B.H. X.W., K.L and M.L. wrote the manuscript with input from all of the authors.

Supplementary Information

# A quantum plasmonic nanocircuit on a semiconductor platform

Xiaofei Wu[1,2,3,4], Ping Jiang[1,5], Gary Razinskas[3,4], Yongheng Huo[6,8], Hongyi Zhang[2,9], Martin Kamp[4], Armando Rastelli[6,10], Oliver G. Schmidt[6], Bert Hecht[3,4], Klas Lindfors[2,7]* and Markus Lippitz[1,2]*

[1]Experimental Physics III, University of Bayreuth, Universitätsstraße 30, 95447 Bayreuth, Germany

[2]Max-Planck-Institute for Solid State Research, Heisenbergstraße 1, 70569 Stuttgart, Germany

[3]Experimental Physics V, University of Würzburg, Am Hubland, 97074 Würzburg, Germany

[4]Wilhelm Conrad Röntgen Research Center for Complex Material Systems, University of Würzburg, Am Hubland, 97074 Würzburg, Germany

[5]College of Science, China University of Petroleum, Changjiang West Road 66, Qingdao 266580, China

[6]Institute for Integrative Nanosciences, IFW Dresden, Helmholtzstraße 20, 01069 Dresden, Germany

[7]Department of Chemistry, University of Cologne, Luxemburger Straße 116, 50939 Köln, Germany

[8]Hefei National Laboratory for Physical Sciences at Microscale and Department of Modern Physics, Shanghai Branch, University of Science and Technology of China, Shanghai 201315, China

[9]Center for Quantum Information, Institute for Interdisciplinary Information Sciences, Tsinghua University, Beijing 100084, China

[10]Institute of Semiconductor and Solid State Physics, Johannes Kepler University Linz, Altenbergerstraße 69, 4040 Linz, Austria

*e-mail: klas.lindfors@uni-koeln.de; markus.lippitz@uni-bayreuth.de

**Dimensions of the designed structure and the coordinate system**

The dimensions of the structure in the design are illustrated in Figure S1. The deviations of the geometry of the real structures from the design are pointed out below in the "Sample fabrications" section.

The coordinate system in Figure S1 is consistent with the ones in the main text and is applied throughout the Supplementary Information.

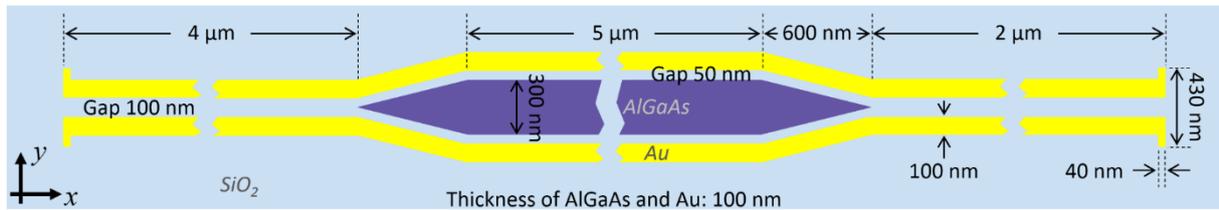

**Figure S1.** Dimensions of the designed structure and the coordinate system (right-handed Cartesian coordinate system).

**Mode analysis of the structure**

Figure S2 presents the two antisymmetric modes and two symmetric modes of the hybrid waveguide and the antisymmetric mode of the two gold wires (the definition of symmetry is based on the charge distribution in the structure) for the as-designed structure. Here the substrate is a half-space of $SiO_2$. The complex dielectric function of gold takes the values from [1], and the refractive indices of $SiO_2$ and the semiconductor bar are set as 1.455 and 3.38, respectively. The latter is taken from [2] for $Al_{0.4}Ga_{0.6}As$. For each mode, the intensities shown in Figure S2 are normalized in such a way that all the modes have the same power.

Looking at the field distribution of the two symmetric modes (Figure S2c,d), one can conclude that not only there is no coupling with a y-polarized dipole at the center due to mismatch of symmetry, but also the coupling with a x-polarized dipole is quite bad, as $E_x$ is still a minor component in these modes. Indeed, the coupling efficiencies with an x-polarized dipole at the center of the AlGaAs are 7.8% and 3.7% (in one propagation direction) for the modes in Figures S1c and S1d, respectively. Therefore, the two symmetric modes are not good for exciting the structure and only the antisymmetric modes are considered in this work.

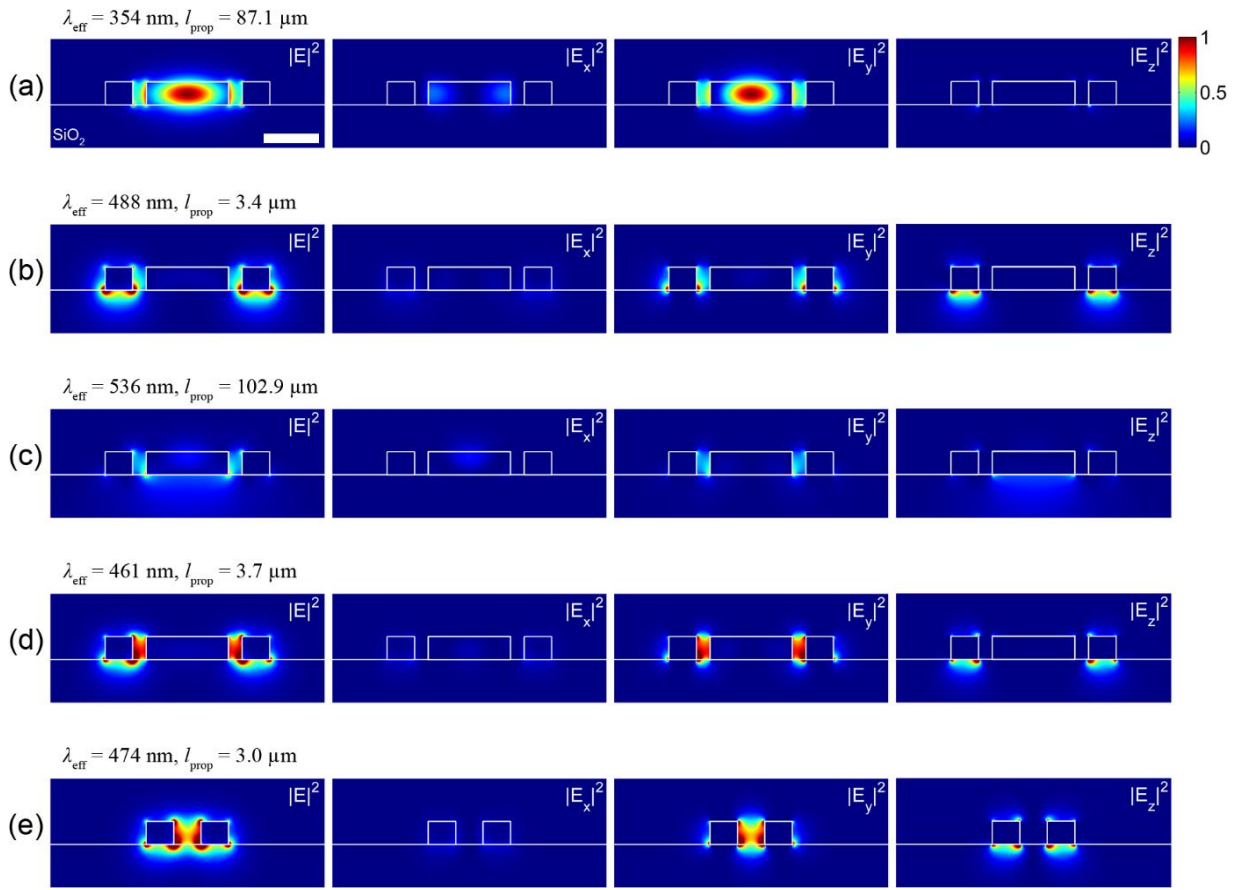

**Figure S2.** Electric field intensity of the antisymmetric modes (a and b) and symmetric modes (c and d) of the hybrid waveguide and the antisymmetric mode of the two gold wires (e) overlaid with the structure profiles (white lines). The effective wavelength ($\lambda_{eff}$) and propagation length ($l_{prop}$) of each mode are shown on the top of each panel. They are derived from $\lambda_{eff} = 2\pi / k_x'$ and $l_{prop} = 1 / (2k_x'')$, where $k_x = k_x' + ik_x''$ is the x-component of the wave vector of each mode. The scale bar in (a) is 200 nm and applies to all panels, as well as the colour scale bar.

# Mode evolution through the tapered part of the structure

Figure S3 exhibits how the two antisymmetric modes of the hybrid waveguide (modes **a** and **b** in Figure S2) evolve when the width of the AlGaAs bar decreases. It is seen that mode **a** exists for every width of the AlGaAs and it gradually evolves to the antisymmetric mode of the gold-wire transmission line (mode **e** in Figure S2) when the AlGaAs vanishes in the end. For mode **b**, however, there is a cut-off width of the AlGaAs. As the width of the AlGaAs decreases, more and more field expands into the substrate. At the point where the effective wavelength of mode **b** is equal to the wavelength in the substrate, the field leaks into the substrate and a well-defined mode **b** that localizes around the waveguide does not exist anymore. As a consequence, the conversion from mode **b** to mode **e** through the tapered part is not as good as from mode **a** to mode **e**.

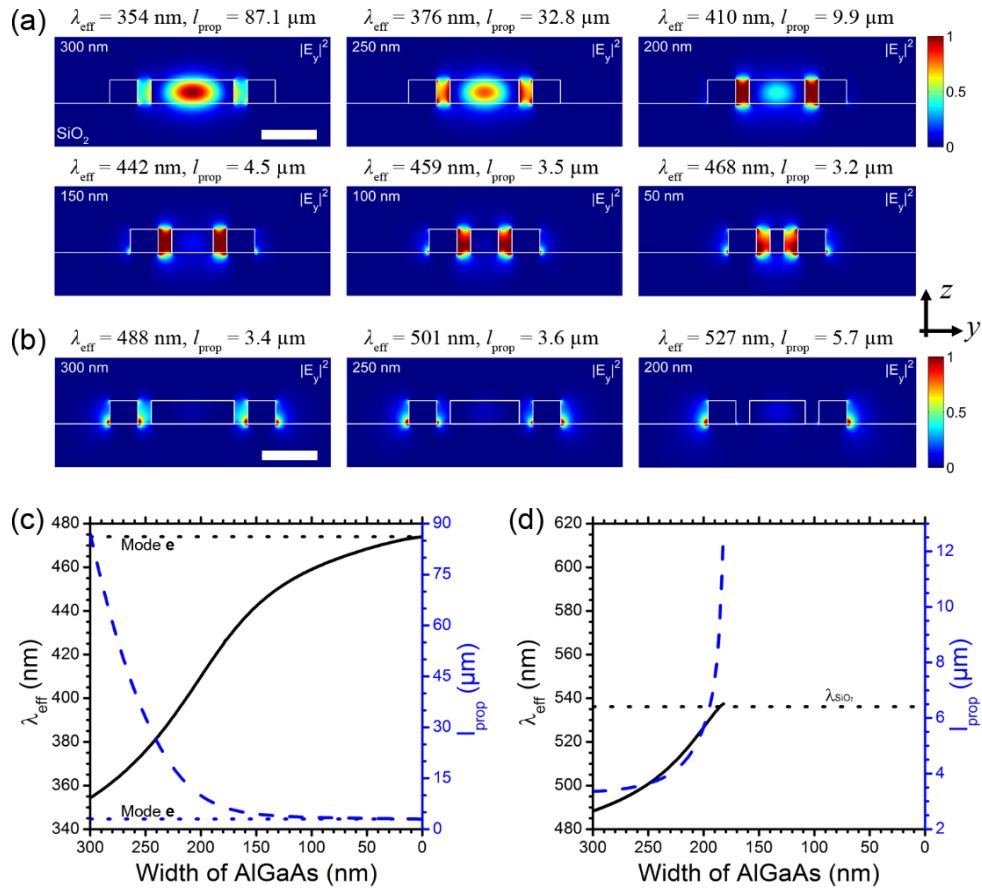

**Figure S3.** Evolution of the antisymmetric modes of the hybrid waveguide with decreasing width of the AlGaAs bar. (a) Intensity of $E_y$ component of mode **a** for different width of AlGaAs (300, 250, …, 50 nm). The effective wavelength ($\lambda_{eff}$) and propagation length ($l_{prop}$) are shown on the top of each panel. (b) Same as the top three panels of (a) but for mode **b**. (c) $\lambda_{eff}$ (black solid line) and $l_{prop}$ (blue dashed line) of mode **a** as a function of the width of the AlGaAs bar. The dotted lines indicate $\lambda_{eff}$ (black) and $l_{prop}$ (blue) of mode **e**. (d) Same as (c) but for mode **b**. The black dotted line indicates the wavelength in the SiO$_2$ substrate for vacuum wavelength of 780 nm. Scale bars: 200 nm.

**Plasmon excitation efficiencies of the two antisymmetric modes of the hybrid waveguide**

The dipole coupling efficiency (α) and mode conversion efficiency through the taper (η) of the two antisymmetric modes of the hybrid waveguide (modes **a** and **b** in Figure S2) for the as-designed structure are derived from simulation. A dipole source oriented in y-direction is placed at the very center of the cross section of the hybrid waveguide. α is defined as the fraction of the source power (set power of the dipole source) coupled into mode **a** or **b**. η is defined as the fraction of the power of mode **a** or **b** converted into the antisymmetric mode of the gold-wire transmission line (mode **e** in Figure S2) through the taper. A closer study on the simulation results reveals that the taper scatters quite some photons into the far-field, which restricts the conversion efficiency, but might be overcome by improved structures. The plasmon excitation efficiency (β) via mode **a** or **b** individually is thus the product of α, η and the exponential decay along the hybrid waveguide. When the dipole is placed 1.1 µm away from the taper tip (Figure 1b in main text), α, η and β are 44.1%, 57.7% and 25.3% for mode **a** and 2.9%, 25.7% and 0.64% for mode **b**, respectively.

Although mode **a** is dominant, mode **b** still plays a role through interference with mode **a**. Figure S4a shows the beating effect on the total plasmon excitation efficiency caused by the co-existence of the two modes. When the dipole is placed properly, mode **b** constructively interferes with mode **a**, resulting in increased efficiencies. Nevertheless, the beating could cause complexity in circuit designs, so sometimes it is still preferred to have only a single mode.

Figure S4b shows that the dipole coupling efficiency of mode **b** decreases quickly as the width of the gold wires increases. It drops to less than 0.1% for gold-wire width of 5 µm. In contrast, the dipole coupling efficiency of mode **a** increases slightly with the gold-wire width. The mode conversion efficiency through the taper exhibits the same trends for modes **a** and **b**, respectively. These results indicate that the beating effect could be significantly suppressed by using wide gold wires.

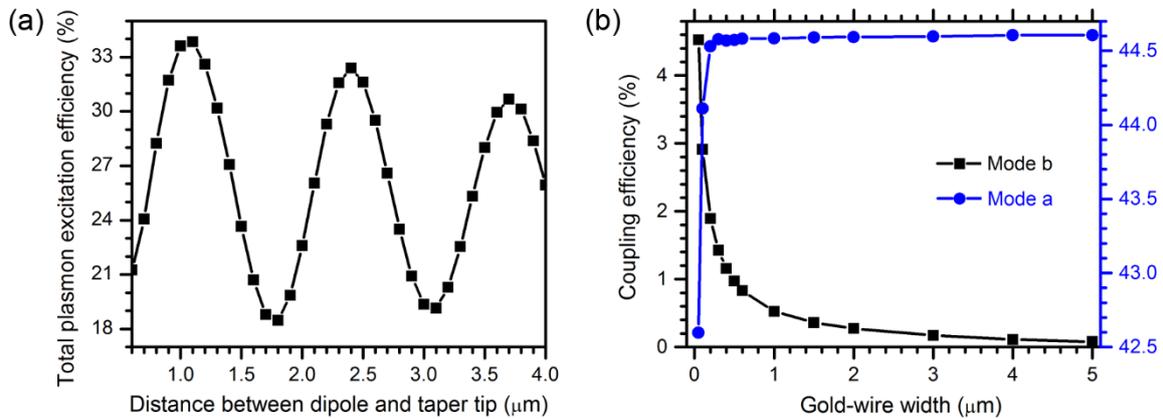

**Figure S4.** Influence of mode **b** to the plasmon excitation efficiency. (a) Beating effect of mode **a** and **b** on the total plasmon excitation efficiency. (b) Variation of the dipole coupling efficiency of mode **a** and **b** with the width of the gold wires.

**Sample fabrications**

A schematic of the self-assembled quantum dot (QD) sample grown by molecular beam epitaxy (MBE) is presented in Figure S5a. The QDs are formed by Al-droplet etching of al AlGaAs layer followed by GaAs filling and AlGaAs overgrowth [3]. The nanoholes are about 10 nm deep. Therefore, the QDs reside at the middle of the layers above the AlAs sacrificial layer, since the top AlGaAs barrier layer is 10 nm less than the bottom one.

An epitaxial lift-off method was used to prepare QD membranes on new substrates following the protocol displayed in Figure S5b. There the membrane in the bottom-left picture was used for producing the sample investigated in this paper. It is remarkable that a whole membrane can be transferred almost completely using such a simple method. Obviously, the large and continuous area facilitates the following fabrication steps on the membranes. The cracks on the membrane are usual due to its fragile nature, but they hardly harm the micro structures made from the membrane.

The scanning electron microscopy (SEM) pictures in Figure S5c show overviews as well as closer inspections of the fabricated structures. The structures look quite good in general. The AlGaAs bar and the gold wires were very well aligned. The AlGaAs bar looks very smooth, whereas the gold wires are somewhat rough, which is usual for evaporated gold due to crystal grains. The optical antennas are in good shape as well.

The exact cross section of the AlGaAs bar was examined with transmission electron microscopy (TEM). The structure in Figure S5d was from a counter-part sample of the real sample (the one used for the optical measurements). Both samples were produced in the same runs of lithography, reactive ion etch (RIE) and atomic layer deposition (ALD). Therefore, the AlGaAs bars should be the same on both samples. However, the second lithography step was not implemented on the counter-part sample, so there is no gold wire on it. It is seen that both sides of the AlGaAs bar are quite steep, forming a nearly rectangular cross section, but the bar is covered with the residual HSQ resist and $SiO_2$ redeposition, which differs from the design. The thickness of the gold wires of the real sample was measured by atomic force microscope, and because of the error in film thickness measurement in the evaporation machine, the gold wires are 10 nm thinner than the design (75 instead of 85 nm). The other dimensions of the real structures, such as the width of the gold wires, the gaps, the antennas and the lengths of the gold-wire transmission lines were characterized by SEM, and they are the same as the designed values.

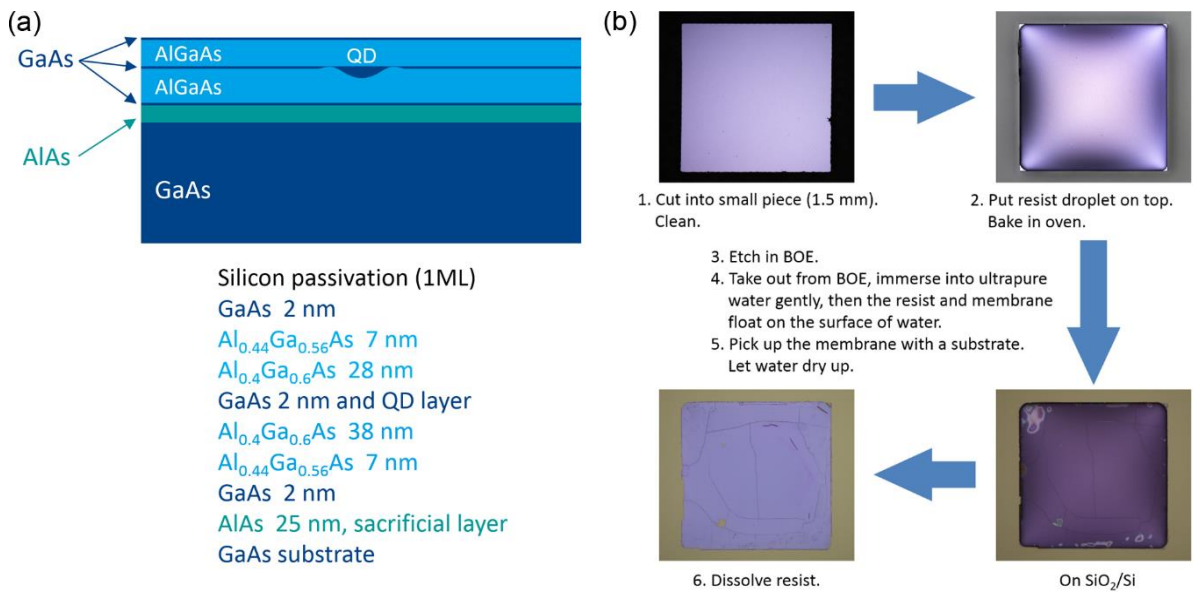

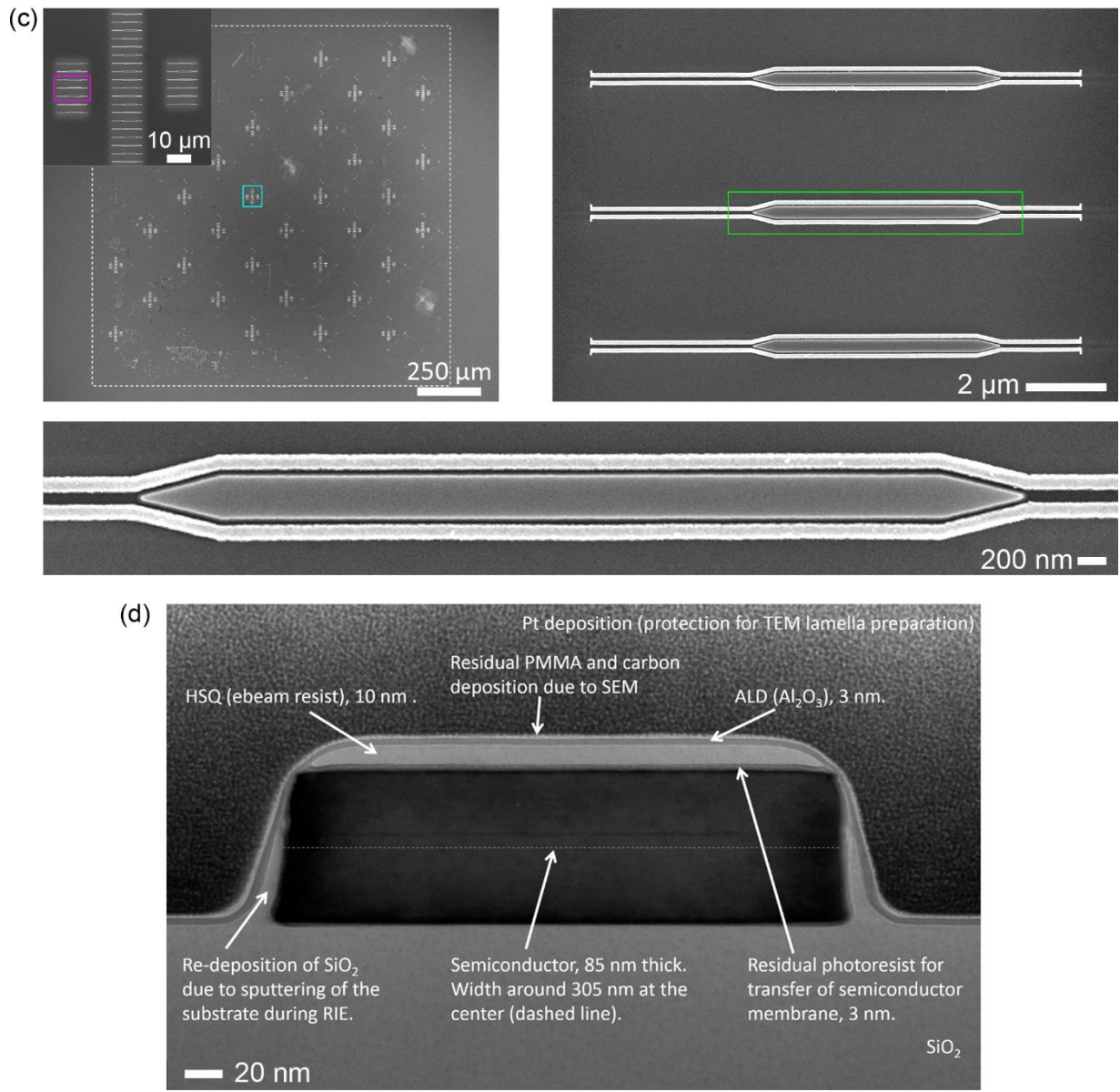

**Figure S5.** Detailed information about the experimental sample. (a) Schematic of a self-assembled GaAs quantum dot in the MBE-grown material, with the specific layers of the material illustrated below. The font colours correspond to the respective layers in the schematic. (b) Steps to produce a QD membrane on a new substrate. (c) SEM pictures of the fabricated structures. Top-left: overviews of the structure arrays. The dashed rectangle indicates the edge of the QD membrane. The inset is an enlarged image of the array in the cyan box. Top-right: SEM image of three structures in the magenta box on the left. Bottom: enlargement of the hybrid waveguide part in the green box. (d) TEM picture of the cross section of a fabricated AlGaAs bar. The slightly darker line above the dashed line is the 2 nm GaAs quantum well layer between the AlGaAs barriers.

**Simulation results for geometry mimicking the real structures**

To see the influence of the deviation of the geometry to the performance of the structure, simulations were implemented for structures with a geometry modified according to the real structures shown in Figure S5. Here the residual PMMA and photoresist layers and the carbon deposition are neglected since they are very thin and their dielectric functions are not well defined. The HSQ on top and the $SiO_2$ deposition on both sides of the AlGaAs are added with the shapes following the reality. Then there is the 3 nm $Al_2O_3$ layer. The thickness of the gold wires is reduced to 75 nm, and the substrate now is 400 nm $SiO_2$ on Si. The refractive indices are 1.41 and 1.635 for HSQ [4] and $Al_2O_3$ [5], respectively. The complex dielectric function of Si takes the values from [6].

From the results in Figure S6, we conclude that the modified geometry does not lead to a pronounced influence to the properties of the structure, such that the structure still performs nearly the same as the as-designed structure in Figure 1 of the main text. Indeed, the dipole coupling efficiency, mode conversion efficiency through the taper and the total plasmon excitation efficiency are 42.8%, 60.4% and 25.6% for the first and 3.4%, 23.0% and 0.66% for the second antisymmetric mode in Figure S6a, respectively. The propagation loss in the hybrid waveguide from the dipole to the taper (0.5 µm) is taken into account for the total plasmon excitation efficiency.

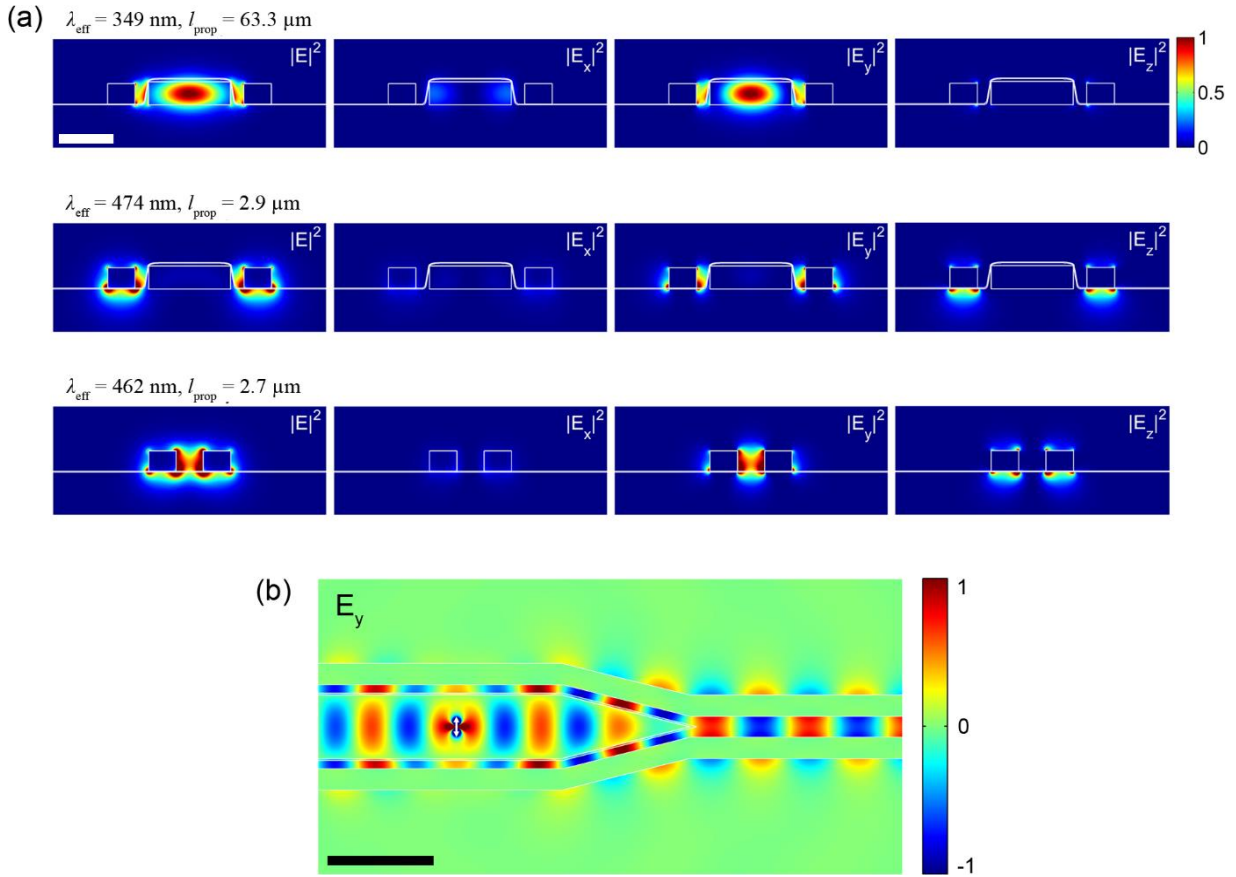

**Figure S6.** Simulations results for geometry mimicking the real structures. (a) Electric field intensity of antisymmetric modes (top and middle) of the hybrid waveguide and the antisymmetric mode of the two gold wires (bottom) overlaid with the structure profiles (white lines). The effective wavelength ($\lambda_{eff}$) and propagation length ($l_{prop}$) of each mode are shown on the top of each panel. The scale bar is 200 nm and applies to all panels, and the colour bar as well. (b) Instant distribution of $E_y$ in the midplane of the AlGaAs bar when the structure is excited by a y-polarized dipole (white arrow). The distance between the dipole and the taper tip is 1.1 µm. The scale bar is 500 nm.

**Unexpected loss in the hybrid waveguide**

In the structure in Figure 2 in the main text, there is actually another quantum dot on the left side (the side of the 4-µm gold-wire transmission line) of the AlGaAs bar. This quantum dot does not exhibit a narrow exciton line but its luminescence still couples into the structure and is radiated out by the antennas eventually. Figure S7 shows a photoluminescence micrograph of the structure when the left quantum dot is excited. It is seen that the left antenna now is brighter than the right one, although the left gold-wire transmission line is twice as long as the right one. Similar results are observed for other structures where the quantum dot is closer to the 4-µm gold-wire transmission line.

The influence of the second antisymmetric hybrid mode (Figure S4a) makes it difficult to evaluate the propagation lengths of the hybrid mode and plasmonic mode, but Figure S7 hints that the propagation loss in the hybrid waveguide is much larger than expected from the simulations. Additional simulations show that a slight deviation (order of $10^{-3}$) of the imaginary part of the refractive index of AlGaAs from zero will drastically increase the loss of the hybrid waveguide. Therefore this unexpected loss could be attributed to the real AlGaAs material that has a slightly different permittivity than the value from literature [2], and could be avoided by moving to longer transition wavelengths [2] or different materials.

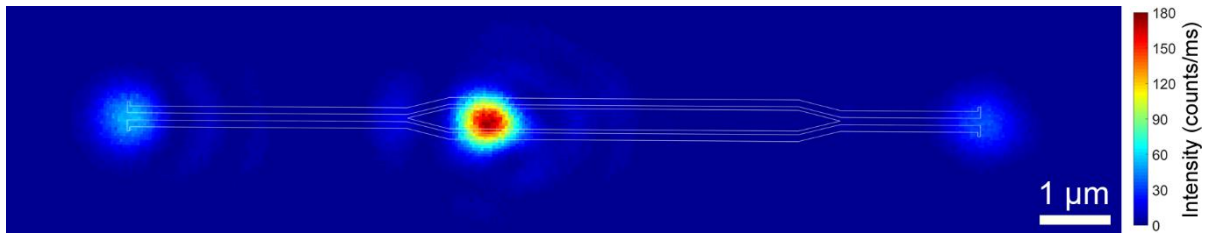

**Figure S7.** Photoluminescence micrograph of the structure at T = 10 K when another quantum dot in the structure is excited by a stationary laser focus (532 nm wavelength). The y-polarized emission is collected by raster-scanning a confocal detection focus.

**An example of a complex integrated quantum nanocircuit**

Figure S8 shows a proposed construction of a fully on-chip quantum nanocircuit using our approach in this work. The circuit integrates three self-assembled QDs. QD1 and QD2 can be electrically excited ($V_1$ and $V_2$) and also tuned ($V_3$ and $V_4$) to match the resonance with QD3, which couples with QD1 and QD2 through a plasmonic splitter [7], constituting a single-photon transistor [8]. Each arm of the splitter can be electrothermally modulated ($V_5$ and $V_6$) to tune the propagating mode between antisymmetric and symmetric modes of the two-wire transmission line [9]. The antisymmetric mode will interact with the QDs but the symmetric mode will just bypass the QDs. The coupling strength of QD1 and QD2 with QD3 can thus be controlled. The output of the transistor is then detected with the on-chip single-plasmon detector [10].

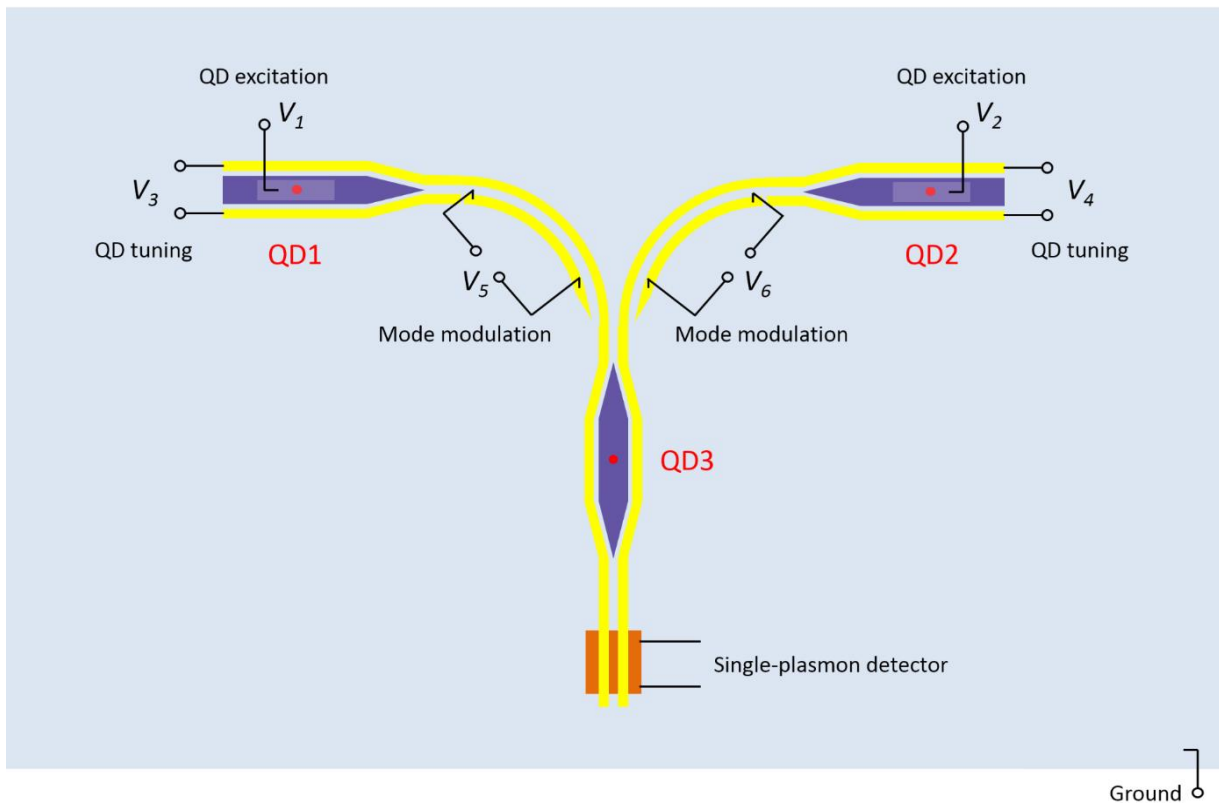

**Figure S8.** Sketch of an all-on-chip single-photon transistor consisting of three self-assembled quantum dots and plasmonic components.

**Optical setup**

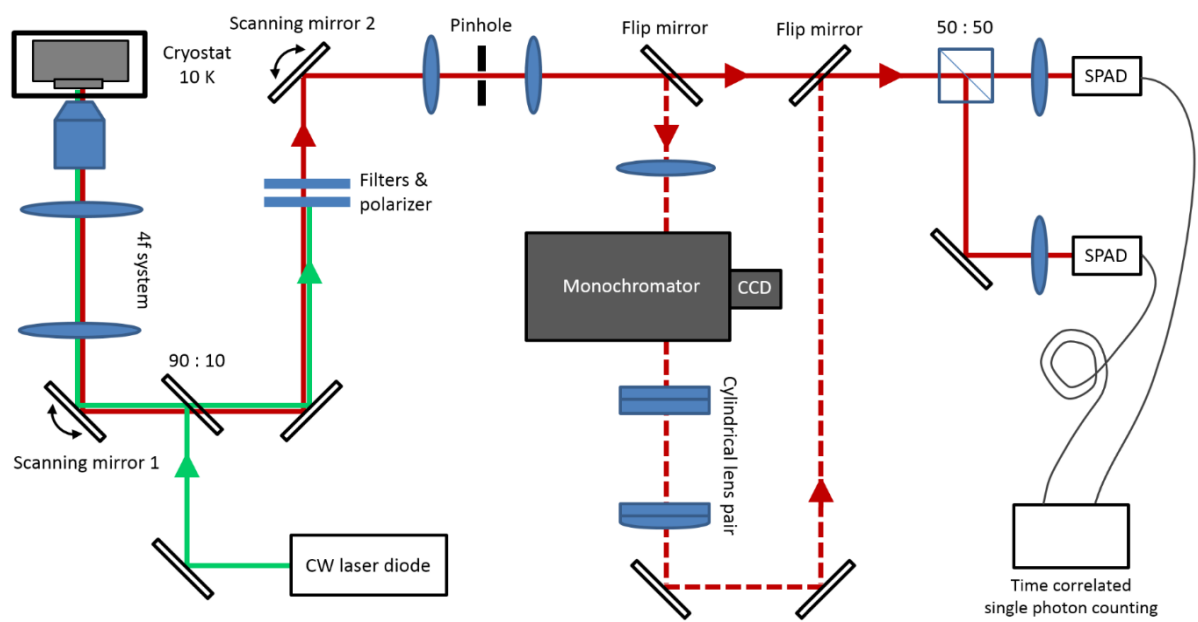

**Figure S9.** Sketch of the optical setup.

**Determination of the position of the quantum dot in the structure**

Our device works the best when the QD is at the center of the structure in the transversal direction (perpendicular to the propagation direction), as the QD coupling efficiency of the antisymmetric modes of the hybrid waveguide drops with the offset from center, whereas the efficiency of the symmetric modes rises. Therefore, the exact position of the QD with respect to the structure is one of the important characteristics of the structure.

Figure S10 shows an example of how the QD position in the transversal direction is characterized for the structure that provided the optical results in the main text. First, as the sample was not perfectly aligned with the optical setup, the original SPAD image has to be rotated by an angle that varies continuously in a certain range, and then the two antenna spots are fitted with a 2-D Gaussian function to determine the center of the spots for each rotation angle. In this way, the rotation angle that makes the line passing the centers of the two spots (i.e. the central line of the structure) horizontal is fixed. Second, under this rotation angle, the spot of the QD is fitted with a 2-D Gaussian function to determine the center of the spot, which indicates the center of the QD. The offset of the QD from the central line of the structure can thus be derived. In this example, the offset that comes out of the fittings is well below 1 nm. Although there is non-zero errors, this result is a sufficient evidence that the QD resides at the very center of this specific structure.

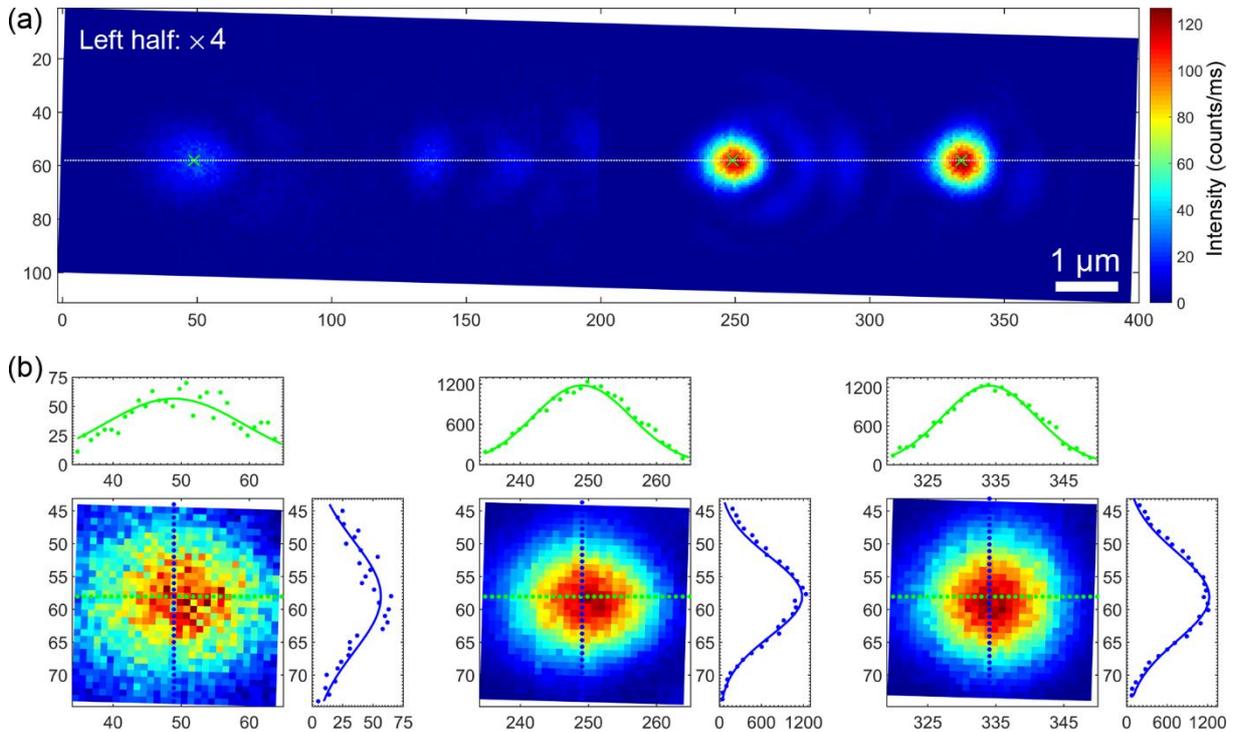

**Figure S10.** Determination of the position of the QD in the structure. (a) The final fitting result. The SPAD image is rotated by 1.6°. The left half of the image is rescaled to make it more visible. The green crosses denote the fitted center of each spot. The dashed line passes the left and right crosses and indicates the central line of the structure. (b) Fitting results for each spot. In each panel, the discrete dots in the top and right plots are the experimental data interpolated along the horizontal (green) and vertical (blue) dotted lines that cross at the fitted center of the spot in the left-bottom graph, whereas the solid curves in the plots are cross sections of the Gaussian function through the green and blue dotted lines, respectively.

**Supplementary Method**

**Modelling photon antibunching for off-resonant excitation.** When using off-resonant excitation, the laser first creates electrons and holes in the local environment of the quantum dot, which then fall into the lower states of the dot and form the exciton. In this way, the single exciton state cannot only be reached from the global ground state (no excited carriers), but also from the states which contain one or more electron-hole pairs in the environment of the quantum dot. The model by Regelman et al. (Ref. 30 in the main text) takes this into account. It sets up a multi-state model with a series of decay channels between these states for InGaAs quantum dots. We adjust the decay rates to our GaAs quantum dots by scaling all rates with a constant factor to fit our experimental data. We solve the system of differential equations using the ground state as start condition. The autocorrelation trace is proportional to the population of the single exciton state, as we detect emission from this state. This leads to the slight bunching around $\tau = \pm 2$ ns. For $g^{(2)}(\tau = 0)$, the model only takes the dark count rate and the temporal resolution of the detector into account. Variation in the excitation rate leads to $g^{(2)} > 1$ for large delay times.